\documentclass[prd, amsfonts, twocolumn, nofootinbib, showpacs]{revtex4}

\usepackage{graphicx, epsfig}
\usepackage{epstopdf}
\usepackage{color}
\usepackage{mathrsfs}
\usepackage{amsmath}
\usepackage{braket}

\newcommand{\be}{\begin{equation}}
\newcommand{\ee}{\end{equation}}
\newcommand{\bea}{\begin{eqnarray}}
\newcommand{\eea}{\end{eqnarray}}

\newcommand{\gapp}{\mathrel{\raise.3ex\hbox{$>$}\mkern-14mu
              \lower0.6ex\hbox{$\sim$}}}
\newcommand{\lapp}{\mathrel{\raise.3ex\hbox{$<$}\mkern-14mu
              \lower0.6ex\hbox{$\sim$}}}

\begin{document}
\title{Modified hoop conjecture in expanding spacetimes and primordial black hole production in FRW universe }
\author{Anshul Saini, Dejan Stojkovic}
\affiliation{ HEPCOS, Department of Physics, SUNY at Buffalo, Buffalo, NY 14260-1500, USA}


\begin{abstract}
According to a variant of the hoop conjecture, if we localize two particles within the Schwarzschild radius corresponding to their center of mass energy, then a black hole will form. Despite a large body of work on the formation of primordial black holes, so far this conjecture has not been generalized to expanding spacetimes. We derive a formula which gives the distance within which two particles must be localized to give a black hole, and which crucially depends on the expansion rate of the background space. In the limit of a very slow expansion, we recover the flat spacetime case. In the opposite limit of the large expansion rate when the inverse Hubble radius is smaller than the Schwarzschild radius of a ``would be" black hole, the new critical distance between two particles that can make a black hole becomes equal to the particle horizon, which is just a requirement that the particles are in a causal contact. This behavior also nicely illustrates why the Big Bang singularity is not a black hole.
We then use our formula to calculate the number density, energy density and production rate of black holes produced in collisions of particles. We find that though black holes might be numerous at high temperatures, they never dominate over the background radiation below the Planck temperature.
\end{abstract}


\maketitle

\section{Introduction}
The formation of primordial black holes (PBHs) in early universe is one of the most important and very active area of research in cosmology and astrophysics.
There are several well studied mechanism of PBH formation, like large density fluctuations in early universe (either from inflation or from cosmological phase transitions), collapse of cosmic string loops, and collapse of closed domain walls   \cite{Rubin:2000dq,Khlopov:1999ys,Khlopov:2008qy,Carr:2005zd,Jedamzik:1996mr,Stojkovic:2005zh,Stojkovic:2004hz,Bugaev:2011wy}. Their potential role in cosmology is very wide, from providing practical solutions to the cosmological monopole and domain wall problems \cite{Stojkovic:2005zh,Stojkovic:2004hz} to playing important role as potential dark matter candidates and source of primordial gravitational waves \cite{Nakama:2016gzw,Bird:2016dcv,Kuhnel:2017bvu,Clesse:2016ajp,Ricciardone:2016ddg,Dong:2015yjs}. They might even affect stability of the Higgs potential by acting as nucleation seeds for the decay of a metastable vacuum \cite{Burda:2015yfa,Greenwood:2008qp}.
 The energy fraction of the universe in PBHs at their formation is well constrained from various observations \cite{Barrow:1991dn,Carr:2009jm,Green:1997sz,Ricotti:2007au,Tashiro:2008sf,Carr:2003bj,Pani:2013hpa}.

The purpose of the work presented in the current paper is two-fold.  First, we present a novel mechanism for PBHs production -- collisions of energetic particles in early universe. Second, virtually all the work done on this topic utilizes a simplifying assumption that the produced black holes are well described by the Schwarzschild solution. However, the collapsing matter in early universe must be unavoidably affected by an expanding background, making the PBHs production more difficult than in Minkowski space. We developed formalism that effectively takes the expanding background into account, and demonstrate that the expansion rate of the universe can significantly affect the PBHs production.

According to the hoop conjecture \cite{hoop}, if we compress a certain amount of mass/energy within its own Schwarzschild radius, then a horizon and thus a black hole will form. In a variant of the original hoop conjecture, if the impact parameter in collision of two particles is smaller than the Schwarzschild diameter for the given center of mass (COM) energy of these particles, then a black hole with the mass equal to the COM energy will form. This mechanism was widely used in search for mini black holes in collider experiments \cite{Dimopoulos:2001hw,Giddings:2001bu,Dai:2007ki}. While collision of two particles is not indeed an analog of a spherically symmetric collapse, there is a strong evidence that the horizon forms in this case too \cite{Eardley:2002re}.  At low temperatures, this process is highly suppressed due to the lack of highly energetic particles, but at high temperatures it cannot be neglected. In \cite{Dai:2016axz,Dai:2017vsm}, the authors have shown that, in a simple thermodynamic system in flat spacetime,  the number density of black holes ($n_{bh}$)  produced in collisions of particles exceeds the number density of particles ($n_{par}$) at a temperature close to but lower than the Planck temperature. Even at low temperatures this process can significantly modify processes that naturally take very long time, like the Poincare recurrence \cite{Dong:2016kuv}.

While we do not expect that PBHs formed in this way in early universe will dominate the energy density at sub-Planckian temperatures, they can be formed in significant numbers and play important role in cosmology and astrophysics. Moreover, number density of PBHs formed by other mechanisms like collapse of the cosmic string loops and density fluctuations during cosmological phase transitions crucially depend on the unknown details of the underlying physics. In that regard, collisions of particles as a mechanism of PBHs production might be one of the most conservative mechanisms. The number density of produced PBHs uniquely depends only on the temperature and expansion rate of the universe, and the predictions might fail only if the hoop conjecture is somehow not correct.

In this paper we consider a homogeneous system consisting of radiation (for simplicity massless bosons) in the background of the expanding Friedman-Robertson-Walker (FRW) universe. In the expanding spacetime it is harder to trap particles, hence the impact parameter yielding a black hole should be smaller than the corresponding value for the same mass/energy in flat spacetime. Therefore, the hoop conjecture should be modified to take the effect of expansion into account. We first do not impose any constraints on the mass of the produced PBHs, and find that the black hole number and energy densities increase monotonically with temperature as expected.

We then repeat analysis for the case where we restrict the masses of the produced PBHs. The first restriction comes from the fact the physics beyond the Planck scale is unknown, hence mass of a PBH should be limited from below by $M_{pl}$.  The second restriction comes from the requirement that only PBH that can survive should be counted. Since black holes radiate via Hawking radiation, in order to create a survivable black hole the background temperature of the radiation must be higher than the black hole temperature. While these restrictions do not significantly affect the distribution of the black holes black holes at high temperatures, they do radically reduce the $n_{BH}$ and $\rho_{BH}$ at low temperatures.

\section{Modified hoop conjecture in expanding spacetime}

In Minkowski space-time, the center of mass energy (COM) in the collision of any two particles with  momenta $k_1$  and $k_2$ is given by
\be \label{come0}
M  =  \sqrt{2 m^2+ 2E_1 E_2  - k_1 k_2 \cos \theta_{1,2} } ,
\ee
where the relativistic energies of the particles are
\be
{E_1}^2 = {k_1}^2  + m^2 ,
\ee
and
\be
{E_2}^2 = {k_2}^2  + m^2
\ee
while $\theta_{1,2}$ is the angle between the momenta $k_1$  and $k_2$. For simplicity, we will consider massless bosons.
In flat spacetime, if the impact parameter in collision of two particles is smaller than the Schwarzschild diameter for the given
COM energy of these particles, then a black hole with the mass equal to the COM energy will form.

\begin{figure}[htpb]
\begin{center}
  \includegraphics[height=0.30\textwidth,angle=0]{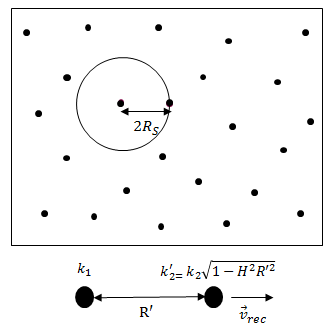}
\caption{In a flat spacetime, if two particle collide with the impact parameter $2R_S$, where $R_S$ is given by the center of mass energy of the colliding particles,  then their collision will lead to the formation of a black hole. However, in an expanding FRW spacetime, particles are moving away from each other with recession velocity, $\vec{v}_{rec}$, which makes it harder to trap them. If a particle with momentum $k_1$ is placed at the coordinate center, then the energy of the particle with momentum $k_2$ will be redshifted by the factor of $\sqrt{-g_{00}}$ in the metric given by Eq.~(\ref{mm}). As a consequence, an impact parameter that yields a black hole will be reduced.}
\label{colfig}
\end{center}
\end{figure}

As we mentioned in the introduction, the expanding background spacetime makes the particles more difficult to trap, so the formation of a black hole should be more difficult than in flat spacetime. Hence the impact parameter must be smaller than $2R_S$, and must depend on the rate of expansion (i.e. Hubble parameter). The FRW metric is
\be
ds^2= - dt^2 + a^2 (t)\left[ d x^2 + x^2 \left( d\theta^2 + \sin^2\theta d\phi^2\right) \right] ,
\ee
where $a(t)$ is the scale factor, and $x$ is the radial coordinate.
It will be useful to write FRW metric in new coordinates related to the comoving coordinates as $\vec{r} = a(t) \vec{x}$. In these coordinates the metric becomes
\be \label{mm}
ds^2 = - (1- H^2 r^2) dt^2 + dr^2 - 2 H r dr dt ,
\ee
where $H \equiv da/dt$ is the Hubble parameter, we work in units where the speed of light $c=1$, and we kept only the radial and temporal part of the metric.  This is the metric in the referent system of one of the colliding particles (in our case the particle with the momentum $k_1$).
To estimate the effect of the expansion on impact parameter, consider two (for simplicity massless) particles at the distance $R'$. The first particle, with momentum $k_1$, is located at the center of the coordinate system, $r=0$, whereas the second particle, with momentum $k_2$, is located at the coordinate $r=R'$, moving away with a recession velocity which depends on its location (see Fig.~(\ref{colfig})). Due to the expansion, the energy of the second particle will be redshifted with respect to the first particle.  The proper time intervals of observers located at $r_1$ and $r_2$ are respectively related to the coordinate time $t$ by
\be
d\tau_1 = \sqrt{-g_{00}(r_1)} dt \, , \ \ \ \ \ \ d\tau_2= \sqrt{-g_{00}(r_2)} dt  .
\ee
Since the change in frequency, $\omega$, is determined by the change in the proper time, by eliminating $dt$ from the above equations we get
\be
\frac{\omega_2}{\omega_1} = \frac{d\tau_1}{d\tau_2} = \frac{\sqrt{-g_{00}(r_1)}}{\sqrt{-g_{00}(r_2)}}  .
\ee
Because all we need is the change in the proper time, the cross term $dr dt$ vanishes, and we recovered the standard redshift which is usually determined by the $g_{00}$ term in the metric.

Another way to see the same thing is to realize that  $-g_{00} = (1- H^2 r^2) = 1- v^2_{rec}$, by using the Hubble's relation $v_{rec} = H r$. In general, the Hubble parameter $H$ can be time dependent, but during the very short duration of the particle collision may be considered to be constant.  Then $\sqrt{-g_{00}}$ becomes just an inverse relativistic gamma factor. Thus, the redshifted energy of the second particle with respect to the first particle is
\be \label{k2p}
k'_2 = k_2\sqrt{-g_{00}}= k_2 \sqrt{1- H^2 R'^2} .
\ee
In Minkowski space-time the center of mass energy is Lorentz invariant, so it is the same in all frames. However, in an expanding universe, this quantity will change with the perceived redshift. In the center of mass frame in the static space-time, the Schwarzschild radius associated with the mass/energy $M$ is $R_S = 2M$ (we work in units where $G=1$). In an expanding universe described by the metric in Eq.~(\ref{mm}), where the particle with $k_1$ is located at the origin, we have  $R'_S = 2 M'$, where $M'$ is the redshifted energy, i.e.
\be
 M' = \sqrt{2 k'_1 k'_2 (1 - \cos\theta_{1,2})} ,
 \ee
while $R'_S$ is the corresponding modified Schwarzschild radius. Since the particle with $k_1$ is located at the origin, its energy is not redshifted, so $k'_1 =k_1$, while $k'_2$ is given by Eq.~(\ref{k2p}).
The strength of gravity determined by the Newton's constant $G$ is the same in both frames, so $M/R_S = M'/R'_S$. However, since any two particles with the center of mass energy $M$ within the distance $R=2R_S$  would make a black hole (as in Fig.(\ref{colfig})), our condition for the impact parameter $R'$ should be $M/(2R_S) = M'/R'$.[Note that $M'$ depends on the distance $r$ through  $k'_2$, so $R' \neq 2R_S'$.] This gives the impact parameter that yields a black hole formation as
\be
R' = 2 R_S \sqrt{\sqrt{1 + 4 H^4 R_S^4} - 2 H^2 R_S^2} .
\label{modradius}
\ee
The relation in Eq.~(\ref{modradius}) represents the modified hoop conjecture for an expanding spacetime. We derived it for the collision of two particles, but since $M$ just represents the total mass/energy in the system, it should be valid in a more general setup of gravitational collapse, perhaps up to some geometric factors of the order of unity.
The only limitation is that we did not take the energy density from the environment into account. This however is not a serious limitation as shown in Appendix A.

Obviously, in the limit of $H \rightarrow 0$, we recover the flat spacetime case as $R' \rightarrow 2 R_S$. In the opposite limit of the large expansion rate, $H \gg R_S^{-1}$, the expression converges to $R' \rightarrow H^{-1}$, which coincides with the particle horizon. This behavior is easy to explain. If $H \gg R_S^{-1}$, then the particle horizon, $H^{-1}$, is smaller than the original Schwarzschild radius, $R_S$, of a``would be" black hole. Then the new critical distance between two particles that can make a black hole, i.e. $R' \rightarrow H^{-1}$, becomes just a requirement that these two particles are in causal contact. Otherwise, even if the particles are inside their own Schwarzschild radius, they still cannot make a black hole since they are not in a causal contact and therefore cannot interact.  This fact also nicely illustrates why the Big Bang singularity is not a black hole despite having infinite energy density. In the extreme limit of $H \rightarrow \infty$  particles are moving away from each other so rapidly that they can not be trapped to form a black hole. In that limit the trapping distance becomes $R' \rightarrow 0$. Of course, the limit $H \rightarrow \infty$ can be taken just for the purpose of illustration, since unknown quantum gravity effects
should get important in that regime.

\section{Calculating black hole number density, energy density, and production rate}

Using the modified hoop conjecture derived in the previous section, we will now calculate the number density, energy density, and production rate of the black holes produced in collisions of particles at a given temperature. We consider a bosonic gas at some temperature $T$. The system is described with the Bose-Einstein statistics, which gives the expected number of particles in the energy state $E_i$  as
\be
f(E_i) = \frac{1}{e^{\frac{E_i}{k_B T}}-1} ,
\ee
where $k_B$ is the Boltzmann constant.
As mentioned in the previous section, the center of mass energy in the collision of any two massless particles with  momenta $k_1$  and $k_2$ is given by
\be \label{come}
M  =  \sqrt{ 2E_1 E_2  - k_1 k_2 \cos \theta_{1,2} } ,
\ee
where the relativistic energies of the particles are
\be
{E_1}^2 = {k_1}^2  ,
\ee
and
\be
{E_2}^2 = {k_2}^2
\ee
while $\theta_{1,2}$ is the angle between the momenta $k_1$  and $k_2$.

Let's single out the particle with momentum $k_1$.  The probability of the collision of this particle with any other particle within the impact parameter $R$  is given by
\be
dP =  \frac{4 \pi}{3} R^3 f(k_2)\frac{ d^3 k_2}{(2\pi)^3}
\ee
According to the hoop conjecture, if $R< R'$, where $R'$ is given by the relation in Eq.~(\ref{modradius}), while $R_S = 2M$  (with $M$ given by Eq.~(\ref{come})), then the collision will yield formation of a black hole. The number density of black holes produced this way, $n_{bh}$,  can be estimated by integrating over the all values of momenta $k_1$ and $k_2$ that both particles can take, i.e.
\be
n_{bh} = \frac{1}{2 (2 \pi)^6} \int \frac{4 \pi}{3} R^3 f(k_1) f(k_2) d^3 k_1 d^3 k_2 ,
\ee
where the volume element is
\be
d^3 \vec{k}_{1,2} = k_{1,2}^2 \sin \theta_{1,2} d\phi_{1,2} dk_{1,2} .
\ee
Integration over $\theta_{1,2}$ is somewhat nontrivial and requires explanation. Since the coordinate system can always be rotated to align the $z$-axis along the particle momentum $k_1$, then the angle $\theta_{1,2}$ will be simply angle $\theta_2$. The integration over $\phi_1$ and $\theta_1$ is straightforward and gives the usual $4 \pi$. Thus, we have
\be
n_{bh} = \frac{1}{12 \pi^3}\int R^3 f(k_1) f(k_2) k_1^2 k_2^2 \sin\theta_2 dk_1 dk_2 d\theta_2 .
\label{nbhexp}
\ee

The nonlinear form of Eq.~(\ref{modradius}) does not allow analytical integration of Eq.~(\ref{nbhexp}) because $k_1$ and $k_2$ are coupled to each other, hence numerical methods must used to evaluate the integral in Eq.~(\ref{nbhexp}). All of the plots shown in this paper are evaluated numerically. However, for illustration purpose, in the approximation of a slow expansion $HR_S<<1$ (which turns out not to be very restrictive because of the smallness of $R_S$), we can proceed analytically. Using this approximation one can write
\bea
R^3 =& (2 R_s)^3(1- 2 H^2{ R_s}^2)^{3/2} \nonumber\\
        \approx& (4 M)^3( 1-  12 H^2 M^2) .
\label{approxrad}
\eea
Obviously, finding an explicit form of the Hubble parameter, $H$, is crucial for our calculations.
 The presence of black holes can modify the usual radiation dominated evolution, so we have to incorporate the contribution from the energy density which is in black holes.

The black hole number density from Eq.(\ref{nbhexp}) can be written in terms of the Hubble parameter  as
\be \label{nbhh}
\begin{aligned}
n_{bh} =  & \frac{1}{12 \pi^3}\int (4M)^3(1-12 H^2 M^2)f(k_1) f(k_2)\\
                &\ \ \ \ k_1^2 k_2^2 \sin\theta_2 dk_1 dk_2 d\theta_2 .
\end{aligned}
\ee
Now $k_1$ and $k_2$ can be separated and integrated analytically. The number density of the radiation particles, $n_{rad}$, is defined as
\bea \label{nrad}
n_{rad} &=&\frac{1}{2 \pi^3}\int f(k) d^3 k \nonumber \\
              &=& \frac{\zeta(3) T^3}{\pi^2} .
\eea
Similarly, $\rho_{bh}$ also depends on $H$ and $T$, and is given by
\bea \label{bhed}
\rho_{bh}& =& \frac{1}{2 (2 \pi)^6} \int M\frac{4 \pi}{3} (R)^3 f(k_1) f(k_2) d^3 k_1 d^3 k_2 \nonumber \\
               & = &\rho_{bh_1} - 12 H^2 \rho_{bh_2}  .
\eea
We assigned here the shorthands $\rho_{bh_1} $ and $\rho_{bh_2}$ as
\bea
\rho_{bh_1}& =&  \frac{64}{3 \pi^3} \int  k_1^4 k_2^4 \chi^2   f(k_1) f(k_2) d k_1 d k_2 d\chi\\
\rho_{bh_2}& =& \frac{128}{3 \pi^3} \int   k_1^5 k_2^5 \chi^3 f(k_1) f(k_2) d k_1 d k_2 d\chi ,
\eea
where $\chi = 1 - \cos \theta_2$. The energy density of radiation, $\rho_{rad}$, depends only on the temperature. In particular,
\be
\rho_{rad} = \frac{1}{(2 \pi)^3}\int k_1 f(k_1)d^3 k_1 ,
\ee
which can be integrated to give
\be
\rho_{rad} = \frac{\pi^2}{30} T^4 .
\ee
Since we know $\rho_{bh}$ and $\rho_{rad}$ as a function of $T$ and $H$, the cosmological Friedmann equation can be used to evaluate $H$ in terms of $T$. The Friedmann equation is
\be
H^2= \frac{8 \pi}{3} (\rho_{rad}+ \rho_{bh}) .
\label{friedmaneq}
\ee
By substituting the expressions for $\rho_{rad}$ and $\rho_{bh}$, $H$ comes out to be
\be
H = \sqrt{\frac{8 \pi}{3}\left[ \frac{\left(\rho_{bh_1} + \frac{\pi^2 T^4}{30}\right)}{32 \pi \rho_{bh_2} +1 }\right]}
\ee
With this explicit form of the Hubble parameter, we can evaluate the black hole number and energy densities in Eqs.~(\ref{nbhh}) and (\ref{bhed}) as a function of temperature.

\begin{figure}[htpb]
\begin{center}
  \includegraphics[height=0.30\textwidth,angle=0]{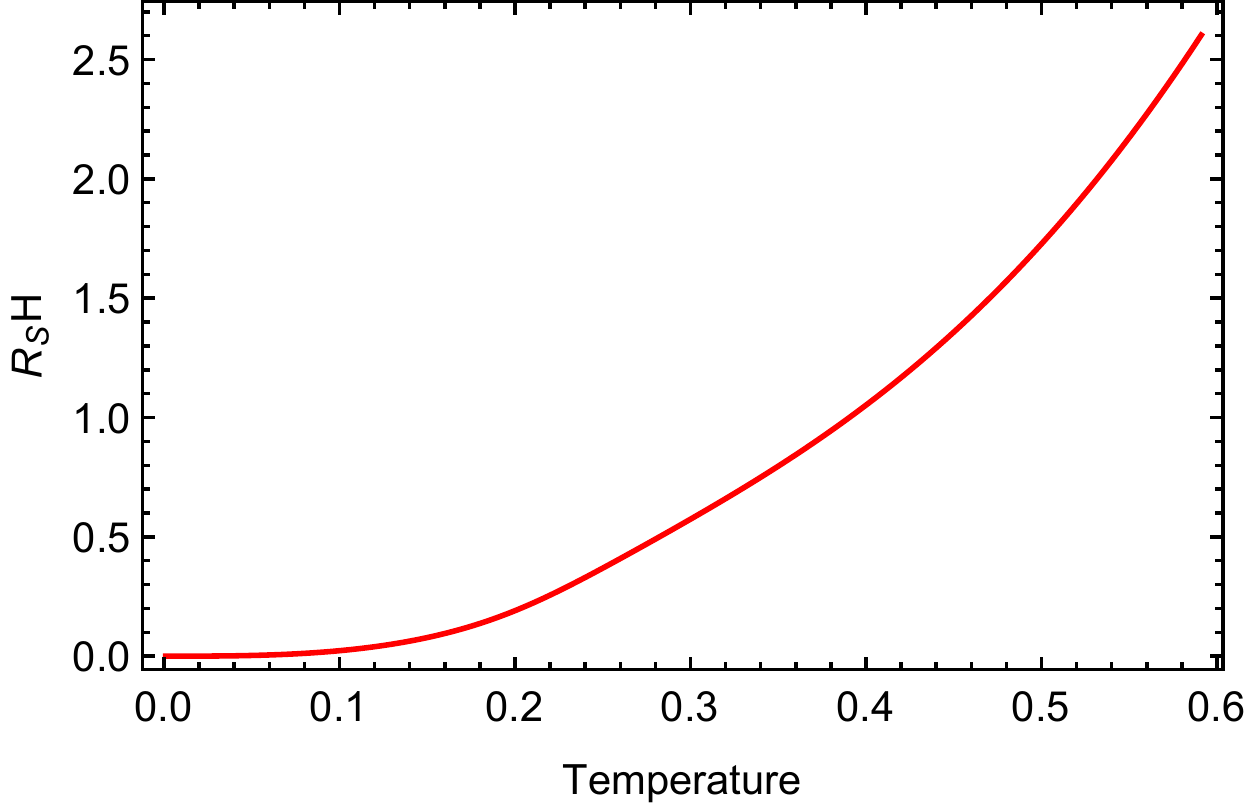}
\caption{We plot $R_S H$ as a function of temperature. It can be seen that the slow expansion approximation ($R_S H <<1$) which enables analytical analysis of the system is valid up to  $T \approx 0.2  T_{pl}$. }
\label{approximation}
\end{center}
\end{figure}

It is now instructive to check the domain of validity of the slow expansion approximation ($R_S H << 1$) in which analytic computations can be done. In Fig.~(\ref{approximation}), we plot $R_S H$ as a function of temperature. Since we used the units in which $G=1$, temperature is measured in units of the Planck temperature $T_{pl}$. We can see that the approximation is valid up to $0.2 \, T_{pl}$ which covers most of the relevant temperature range.

The rate of black hole production ($\dot{n}_{bh} \equiv dn_{bh}/dt$) can be calculated in the similar fashion.
To calculate $\dot{n}_{bh}$, let us take again two particles with momenta $k_1$ and $k_2$.  Locating the particle with momentum $k_1$  at the origin, the number of collisions between this single particle and all other particles within the time interval $dt$ is given by
\be
dN_{col} =\frac{ n_{rad} dt}{(2\pi)^3} \int d^3 k_2 v_{12}(\vec{k}_{1}, \vec{k}_{2})  {\sigma}_{12}(\vec{k}_{1}, \vec{k}_{2}) f(E_2),
\ee
where $v_{12}$ is the relative velocity of the particles, while ${\sigma}_{12}$ is the black hole production cross section.

Then, $\dot{n}_{bh}$ is obtained by integrating $dN_{col}/dt$ over all the values of the momentum $k_1$ as
\be \label{ndot}
\frac{dn_{bh}}{dt} = \frac{1}{2(2\pi)^3} n_{rad} \int d^3 k_1 f(E_1) \frac{d N_{col}^{(1)}}{dt} .
\ee
The above expression may be explicitly written as
\be
\begin{aligned}
\frac{dn_{bh}}{dt} = &\frac{1}{2 (2\pi)^6} n_{rad}^2\int \int d^3 k_1 d^3 k_2  v_{12}(\vec{k}_{1}, \vec{k}_{2})  \\        &{\sigma}_{12}
                (\vec{k}_{1}, \vec{k}_{2})  f(k_1) f(k_2)
\label{nbhrate0}
\end{aligned}
\ee
\\

The peculiar relative velocity between two particles is
\bea
 v^p_{12}(\vec{k}_{1}, \vec{k}_{2}) =&   \mid \vec{ v}_{1} - \vec{ v}_{2} \mid \nonumber \\
 =&  \sqrt{v_1^2 + v_2^2 - 2 v_1 v_2 \cos \theta_2} .
\label{relvel}
\eea
The individual velocities of the particles are defined as $v_1 = \frac{k_1}{E_1}$ and $v_2 = \frac{k_2}{E_2}$. Therefore
\be
 v^p_{12}(\vec{k}_{1}, \vec{k}_{2}) = \sqrt{{\left(\frac{ k_1}{E_1}\right)}^2 + {\left(\frac{ k_2}{E_2}\right)}^2  - 2\frac{k_1 k_2}{E_1 E_2} \cos \theta_2} .
\ee
The expansion of the background spacetime has also an effect on relative velocities of any two particles. The total relative velocity must pick up the contribution from the  recession velocity. Thus, the total relative velocity $v_{12}$ is given by
\bea
 \vec{v}_{12} &=& v^p_{12}(\vec{k}_{1}, \vec{k}_{2}) - \vec{v}_{recession}\\
                     &=& v^p_{12}(\vec{k}_{1}, \vec{k}_{2}) - H R
\label{modvel}
\eea
The black hole production cross section is defined as
\bea
\sigma_{12} &=& \pi R^2 \nonumber \\
                    &\approx&\pi \left[ 4 M \sqrt{1- 8 H^2 M^2}\right]^2 .
\label{modcross}
\eea
Substituting $v_{12}$ and $\sigma_{12}$ from Eq.~(\ref{modvel}) and Eq.~(\ref{modcross})  into Eq.~(\ref{nbhrate0}) gives,
\be
\begin{aligned}
\frac{dn_{bh}}{dt} = &\frac{1}{16 \pi^3} n_{rad}^2\int \int k_1^2 k_2^2 d k_1 d k_2 d \chi \left(\sqrt{2 \chi} - H R \right) \\     & R^2  f(k_1) f(k_2)
\label{nbhrate}
\end{aligned}
\ee
where $\chi = (1- \cos(\theta_2))$,  $n_{rad}$ is the number density of the radiation particles, while the impact parameter, $R$, is given in Eq.~(\ref{modradius}). This is an exact formula without any  approximations.  To simplify it a bit, an approximate expression for the impact  parameter  from Eq.~(\ref{approxrad}) can be used to arrive at the approximate final expression for $\dot{n}_{bh}$  as
\begin{widetext}
\bea
\frac{d n_{bh}}{dt} = \frac{1}{2} \frac{n_{rad}^2}{(2\pi)^6}  \int\int 256 {\pi}^3 {k_1}^3  {k_2}^3 \chi (\sqrt{2 \chi}  (1- 8 H^2 M^2)- 4 H M ) f(k_1)f(k_2) d\chi dk_1 dk_2 ,
\eea
\end{widetext}

\section{Integrating equations with no restriction on the black hole mass}

In this section, we will consider an ideal scenario in which the black hole masses are not restricted, so $k_1, k_2$ and $\theta$ can take any value allowed in their domain. In that case, the integral for the black hole number density in Eq.~(\ref{nbhh}) can be solved exactly to give
\be
n_{bh} = a_1 T^9  - a_2 H^2 T^{11} ,
\label{nbh}
\ee
with $a_1 =\frac{1470  \zeta \left(\frac{9}{2}\right)^2}{\pi ^2} $ and $a_2 =  \frac{1020600  \zeta \left(\frac{11}{2}\right)^2}{\pi ^2}$, where $\zeta$ is the Riemann zeta function. Similarly, black hole energy density in Eq.~(\ref{bhed}) comes out to be
\be
\rho_{bh} = b_1 T^{10}  -  b_2  H^2 T^{12} ,
\label{rhobh}
\ee
 where $b_1  =  \frac{32768 \zeta(5)^2}{\pi^3} $ and $b_2  =   \frac{131072 \pi^9}{3969}$. As expected, the terms containing $H$ have negative contribution in both $n_{bh}$ and $\rho_{rad}$, which means that the expansion is suppressing black hole creation. Also, using Eq.~(\ref{friedmaneq}), $H$ can be written as
\be
H =\sqrt{ \frac{b_1 T^{10} +\left(\frac{\pi^2}{30}\right)T^{4} }{\left(\frac{3}{8 \pi}\right) + b_2 T^{12}}} .
\label{hubble}
\ee

\begin{figure}[htpb]
\begin{center}
  \includegraphics[height=0.30\textwidth,angle=0]{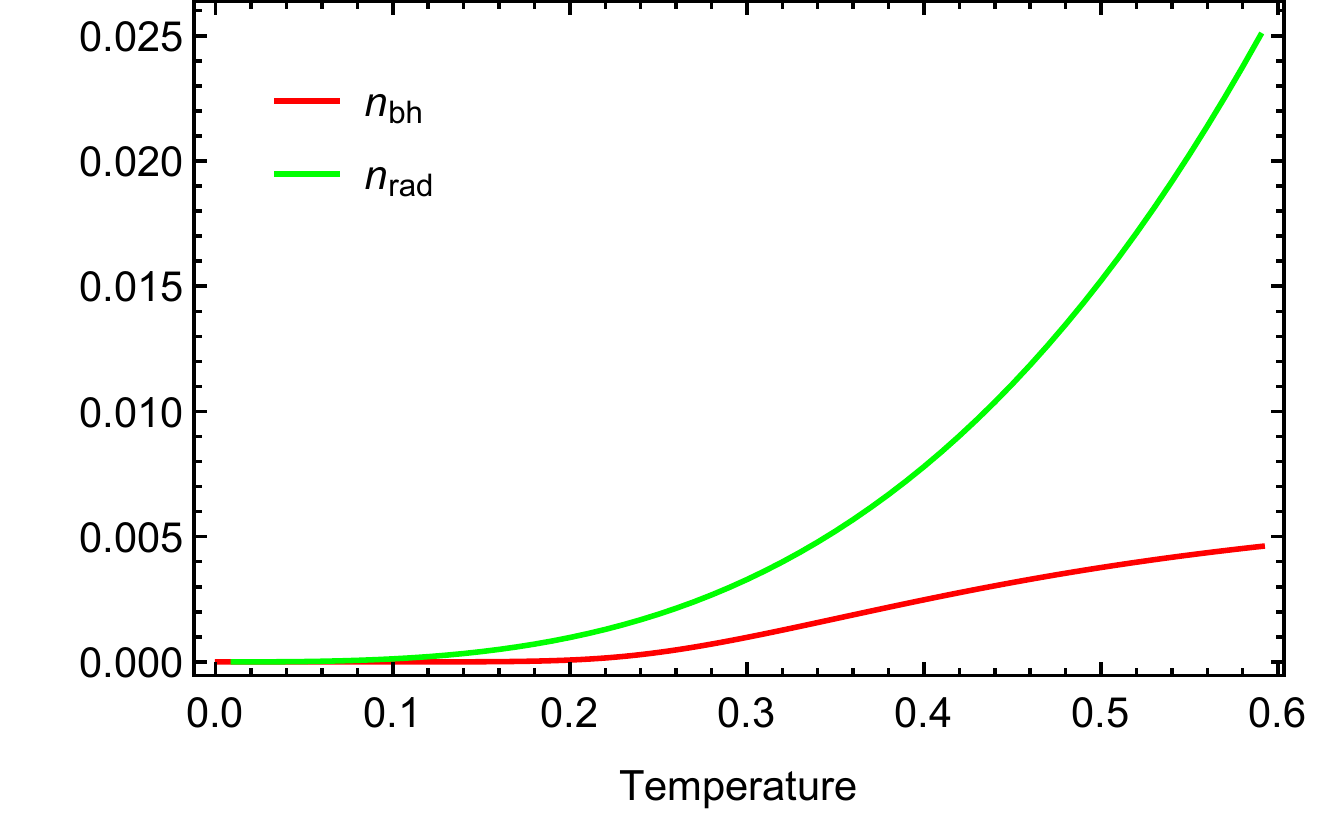}
\caption{We plot black hole number density, $n_{bh}$, and particle number density $n_{rad}$ as a function of temperature. It is clear that black holes do not dominate the system ($n_{rad}$\textgreater $n_{BH}$) in the explored regime. This result is in contrast with the flat space time where $n_{bh}$ dominate at $T> 0.3 \, T_{pl}$. The difference comes from the expansion of universe which makes harder to trap particles.}
\label{nocons_nbhnrad}
\end{center}
\end{figure}

\begin{figure}[htpb]
\begin{center}
  \includegraphics[height=0.30\textwidth,angle=0]{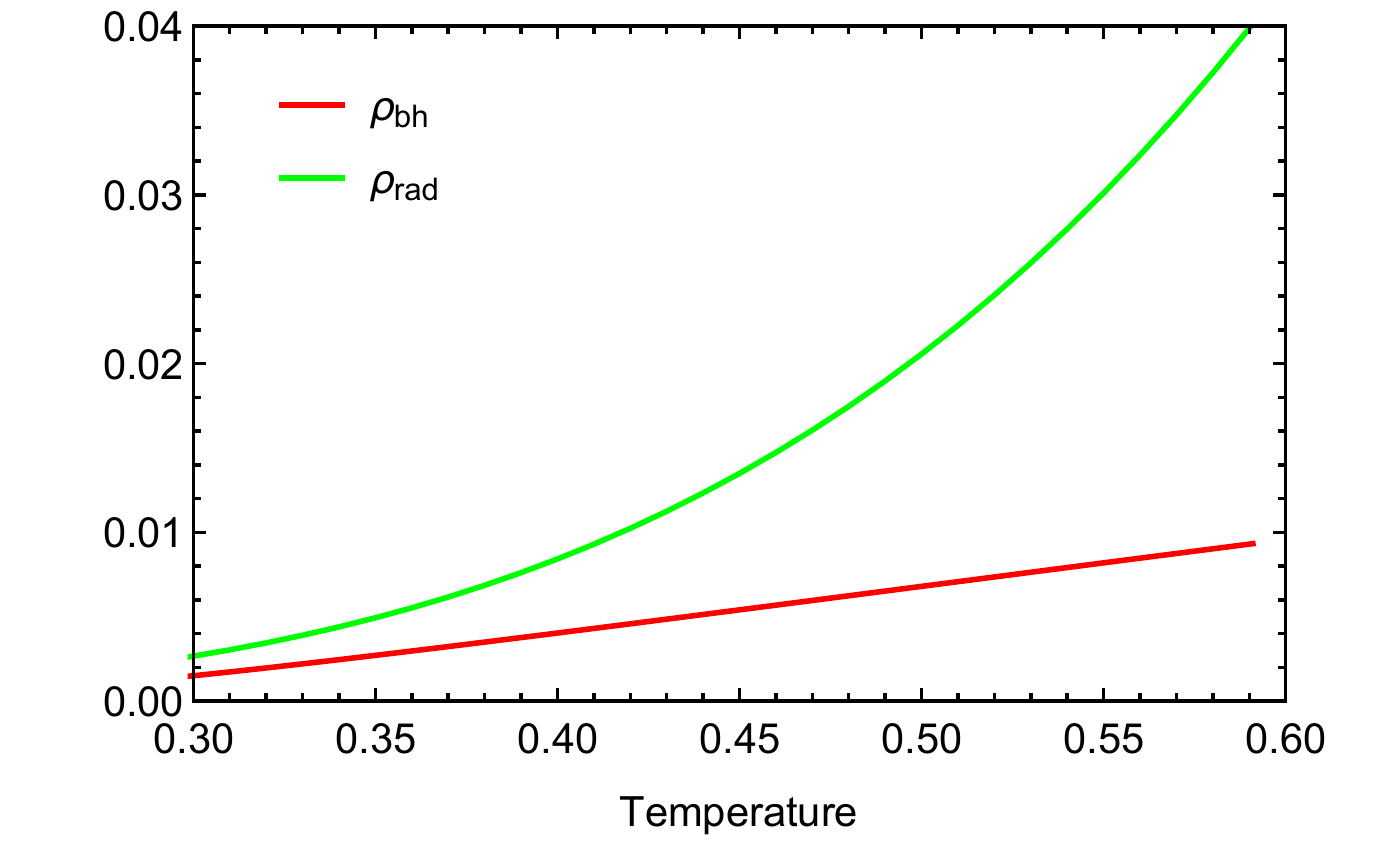}
\caption{We plot black hole energy density, $\rho_{bh}$, and radiation energy density $\rho_{rad}$ as a function of temperature. Energy density of black holes never dominates the system at any temperature.}
\label{nocons_rhobhrhorad}
\end{center}
\end{figure}

High temperature in early universe affects the relevant physics in two ways.  At high temperatures, black hole production in collisions of particles steps up.  On the other hand, the value of the Hubble parameter also grows with temperature, which in turn suppresses production of PBHs. Since these are two competing effects, it is not a priori clear how the system behaves.
To illustrate what is going on precisely, in Fig.~(\ref{nocons_nbhnrad}) we plot $n_{bh}$ and particle number density $n_{rad}$ as a function of temperature. As noted above, all the plots are made from the exact expressions performing numerical calculations. In this case, $n_{bh}$ and $n_{rad}$ are given by Eq.~(\ref{nbhexp}) and Eq.~(\ref{nrad}). It can be seen that $n_{rad}$ dominates over $n_{bh}$ in the whole plotted temperature range, but at high temperatures they are comparable. This means that, while black holes can be very numerous in early universe, they will never dominate the system by their sheer numbers. This is in contrast with flat spacetime where around $T \approx 0.3 \, T_{pl}$ black holes start dominating the system by numbers. Obviously, the difference comes from the effects of expansion, since as temperature increases the Hubble parameter increases too, making formation of PBHs more difficult. Similarly, in Fig.~(\ref{nocons_rhobhrhorad}) we compare the energy densities and show that black holes never dominate.

\begin{figure}[htpb]
\begin{center}
  \includegraphics[height=0.30\textwidth,angle=0]{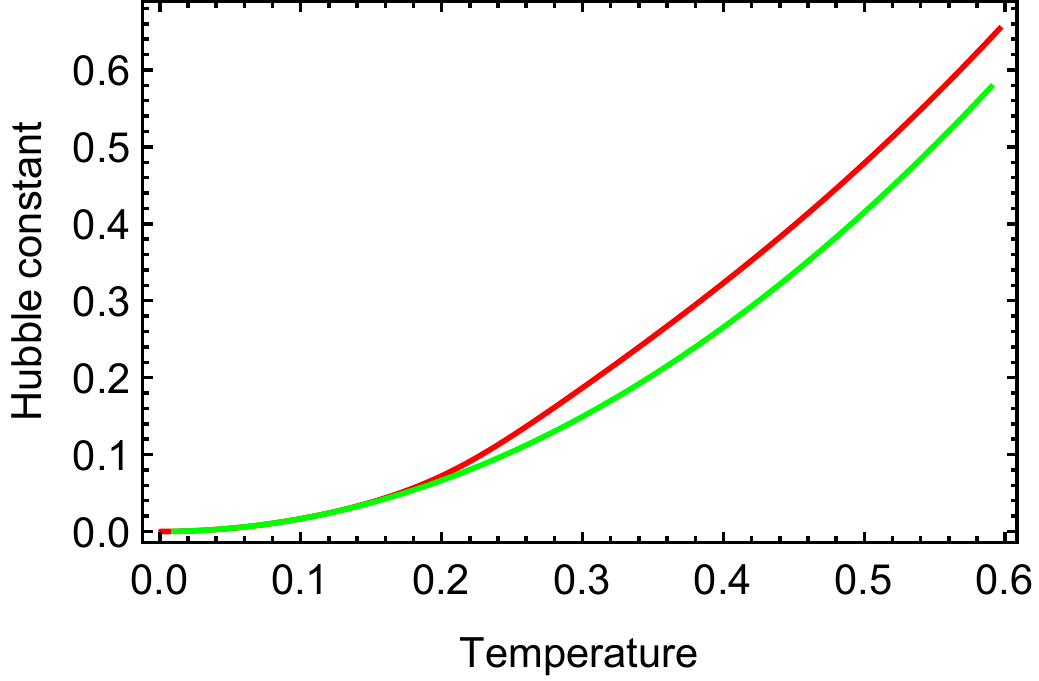}
\caption{We plot the Hubble parameter, $H$, as a function of temperature.  The red curve represents $H$ in the early universe containing black holes, while the green curve is $H$ in the usual radiation dominated universe. At low temperatures they are indistinguishable, however at high temperatures the black hole energy density, $\rho_{bh}$, is significant and contributes to higher values of $H$.}
\label{nocons_hubble}
\end{center}
\end{figure}

In Fig.~(\ref{nocons_hubble}), we plot the Hubble parameter, $H$, a function of temperature. $H$ is calculated exactly from Eq.~(\ref{friedmaneq}), and we can see that it increases monotonically at all temperature as it should. However, compared with the usual radiation dominated $T^2$ dependence, we see that at very high temperatures, $H$ increases faster due to the contribution from $\rho_{bh}$.

\begin{figure}[htpb]
\begin{center}
  \includegraphics[height=0.28\textwidth,angle=0]{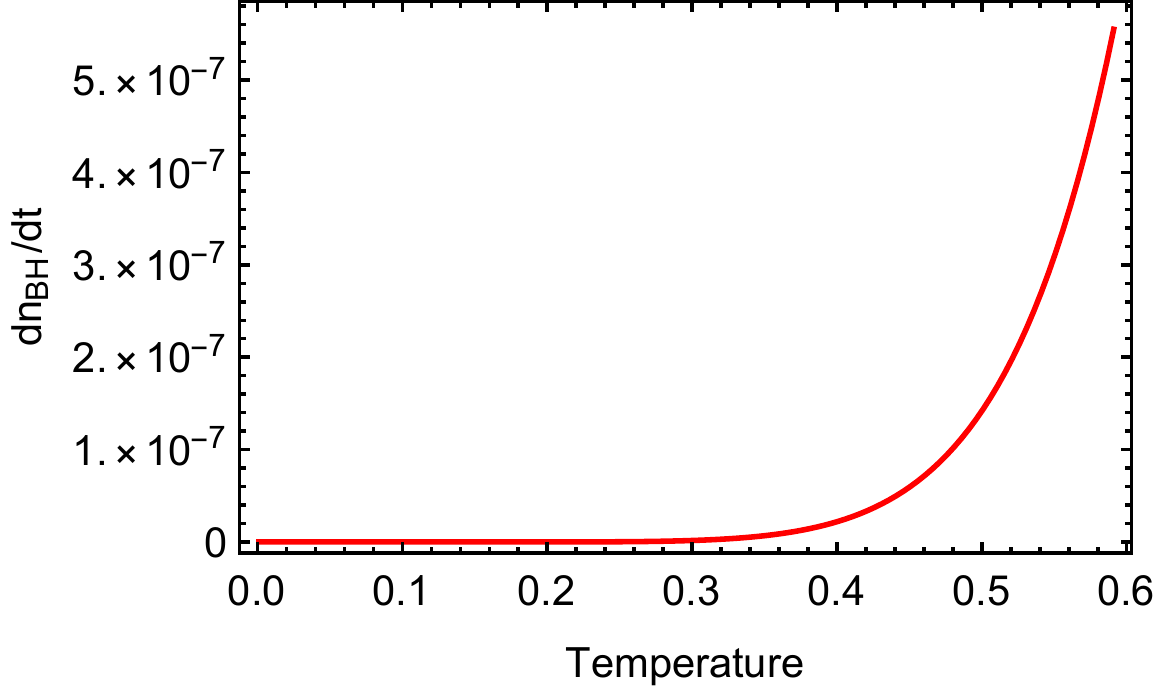}
\caption{We plot the black hole production rate, $\dot{n}_{bh}$, as a function of temperature temperature. As expected the black hole production increases monotonically as temperature increases.}
\label{nocons_nbhrate}
\end{center}
\end{figure}

In Fig.~(\ref{nocons_nbhrate}), we plot the black hole production rate, $\dot{n}_{bh}$, as a function of temperature, using the exact expression from Eq.~(\ref{nbhrate}). The rate increases monotonically with temperature, as expected.

\section{Black holes heavier than the Planck mass and colder than the surrounding radiation}

In previous section we did not put any constraints on the produced black hole mass. However, black holes with sub-Planckian masses (i.e. $M < M_{pl}$) have Hawking temperature higher  than $T_{pl}$, which cannot be reliably described by a semi-classical theory. To avoid this problem, we require that the produced black holes are more massive than $M_{pl}$. Since the produced black hole mass depends on the collision angle $\theta_2$ and momenta $k_1$ and $k_2$ through Eq.~(\ref{come}), the constraint $M > M_{pl}$ restricts the values that the angle $\theta_2$ can take as
\be
\chi > \frac{1}{2 k_1 k_2} ,
\label{massconstraint}
\ee
where $\chi = 1- \cos(\theta_2) $. Since  $0< k_1, k_2 < \infty$, the parameter $x$ can potentially take values larger than $2$, which falls beyond the valid domain of $\chi$ (i.e. $\chi <2 $). To satisfy condition in Eq.~(\ref{massconstraint}), $k_1$ and $k_2 $ must satisfy relation $k_2 > \frac{1}{4 k_1}$. Due to this change in integration limits, the relevant integrals can not be solved exactly even in the slow expansion limit. Therefore, the integration must be done numerically.

\begin{figure}[htpb]
\begin{center}
  \includegraphics[height=0.30\textwidth,angle=0]{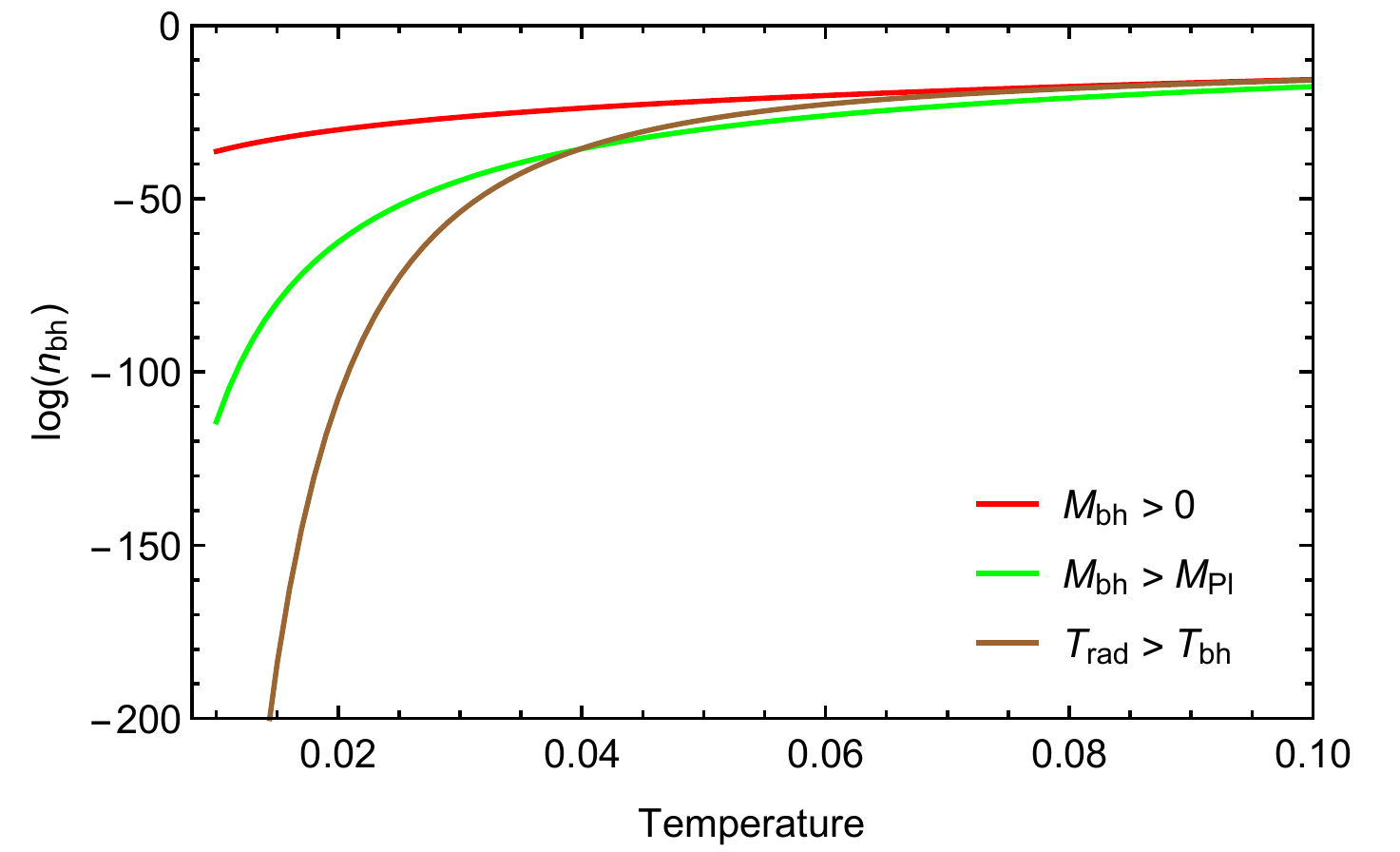}
\caption{We plot $\log(n_{bh})$ as a function of temperature for different constraints imposed on the produced black hole mass. At high temperatures the black hole number density is similar for each case. However at low temperatures the number density decreases most rapidly for black holes colder than the environment, $T_{bh}\textless\   T_{rad}$. At $T \sim 0.04 \, T_{pl}$, there is a crossover above which is easier to create a black hole colder than the environment than a black hole heavier than $M_{pl}$.}
\label{nbhcomparison}
\end{center}
\end{figure}

Another constraint that we want to impose comes form the requirement that we do not want the produced PBHs to disappear right after they are produced. In other words, we want the
Hawking temperature of the produced PBHs, $T = (8 \pi M)^{-1}$, to be lower than the temperature of the surrounding radiation at the moment of formation. For such black holes accretion will dominate evaporation, and PBHs will remain in the system. [To answer how long they can survive requires detailed analysis of the time-dependent coupled system PBHs-radiation, where PBHs and radiation can exchange energy, however, this is outside of the scope of this paper.]
This condition also restricts the mass of black holes and thus the collision angle
\be
\chi > \frac{1}{128 \pi^2 {k_1} {k_2} T^2}
\ee
Again, to satisfy $2 > \chi >0$, we must have $k_2 > \frac{1}{256 \pi^2 k_1 T^2}$, while $k_1$ can take any positive real value.
We plot $\log(n_{bh})$ in Fig.~(\ref{nbhcomparison}) with these constraints taken into account and compare it with the case with no constraints. At high temperatures, $\log(n_{bh})$ is of the same order for every scenario. However at low temperatures, the constraints significantly reduce the number of produced PBHs. One can see that the constraint $T\textgreater T_{bh}$ is more restrictive than $M\textgreater M_{pl}$ at low temperatures. It is also possible to notice a crossover at $T \sim 0.04 \, T_{pl}$, above which is easier to create a black hole colder than the environment than a black hole heavier than $M_{pl}$.
We also note that in the explored temperature range there is no significant change in behavior of the Hubble parameter for all three plotted cases.

\section{Conclusion}
\label{conclusion}
We investigated here the question of primordial black holes production in a thermodynamic system consisting of radiation in an expanding FRW universe. We used a well known fact that the particles collisions can lead to formation of black holes if particles are confined within the Schwarzschild radius corresponding to their center of mass energy. This is known as a variant of the hoop conjecture. However,
the hoop conjecture has originally been formulated for the flat spacetime, and has never been generalized to expanding backgrounds.  In this paper, we first generalized the hoop conjecture to the case of an expanding FRW universe. Our formula crucially depends on the Hubble parameter (i.e. the rate of expansion), and fatefully reproduces the flat spacetime limit. The impact parameter leading to formation of black holes is smaller than the corresponding flat spacetime value. In the limit of a very large Hubble parameter, $H$, (i.e. when the $H^{-1}$ is smaller than the original Schwarzschild radius of a``would be" black hole), the new critical distance between two particles that can make a black hole becomes just a particle horizon $H^{-1}$. This is of course just a requirement that two colliding particles are in a causal contact, but it factors out naturally from our derived formula for the modified hoop conjecture. In the extreme limit of $H \rightarrow \infty$  particles are moving away from each other so rapidly that they can not be trapped to form a black hole. This is a nice illustration of the fact that the Big Bang singularity is not a black hole.

We then applied our formula to calculate the number density of black holes, $n_{bh}$, created in collisions between particles in the hot system of an expanding early FRW universe.
Since the background spacetime is expanding, the probability of formation of black holes due to the particle collisions is reduced.
We first performed analytic calculation without imposing any restriction on the black hole mass, in the slow expansion approximation, $HR_S<<1$, which is not very restrictive due to the smallness of $R_S$. All the relevant quantities were calculated as a function of the temperature of the radiation (the free parameter in the system).  We found that $n_{bh}$ increases monotonically with temperature but it always remain smaller than the radiation number density, $n_{rad}$. This is in contrast with flat space time where $n_{bh}$ dominates around $T \sim 0.3 \, T_{pl}$. Similarly $\rho_{bh}$ also increases monotonically with temperature. We also plotted the product $R_S H$  with  temperature and it remain  less than unity up to $T \approx 0.2 T_{pl }$ which covers most of the relevant temperature range justifying the slow expansion approximation in our analytic calculations.

We then repeated calculations for black holes which are heavier than the Planck mass (semiclassical case), and colder than the surrounding radiation (for which accretion dominates evaporation, and they remain in the system). While at high temperatures, the number of produced black holes does not depend much on the mass restriction, at low temperatures the constraints significantly reduce the number of produced black holes. It is also possible to notice a crossover at $T \sim 0.04 \, T_{pl}$, above which is easier to create a black hole colder than the environment than a black hole heavier than $M_{pl}$.

At the end we mention somewhat related work in \cite{Kleban:2016sqm} and \cite{East:2015ggf} where black hole production in anisotropic cosmologies due to large density perturbations were studied.

\begin{acknowledgments}
This work was partially supported by the NSF grant number PHY-1417317.
\end{acknowledgments}

\appendix
\section{Neglecting the background energy density }

In our derivation of the formula for the modified hoop conjecture in Eq.~(\ref{modradius}), we took into account only the center of mass energy of two colliding particles, and neglected the background energy density of the FRW metric. In other words, the mass of the produced black hole, $M_{bh}$, is equal to the (appropriately redshifted) center of mass energy of the two colliding particles, $M'$, instead of the Misner-Sharp-Hernandez (MSH) mass.  Strictly speaking, a sphere of radius $R'$ has the MSH mass
\be
M_{\rm MSH} = M' + \frac{H^2R'^3}{2}=M' + \frac{4 \pi}{3} \rho R'^3 ,
\ee
Our approximation where we neglected the contribution from $\rho$ should be valid if the background energy density does not gravitate, for example in the case of the de Sitter space with $\rho = \Lambda / (8 \pi)$, where $\Lambda$ is the cosmological constant. In the general case, this approximation is valid if $M' \equiv M_{bh}>> \frac{4 \pi}{3} \rho_{rad} R'^3$, where $R'$ is given by Eq.~(\ref{modradius}).

\begin{figure}[htpb]
\begin{center}
  \includegraphics[height=0.30\textwidth,angle=0]{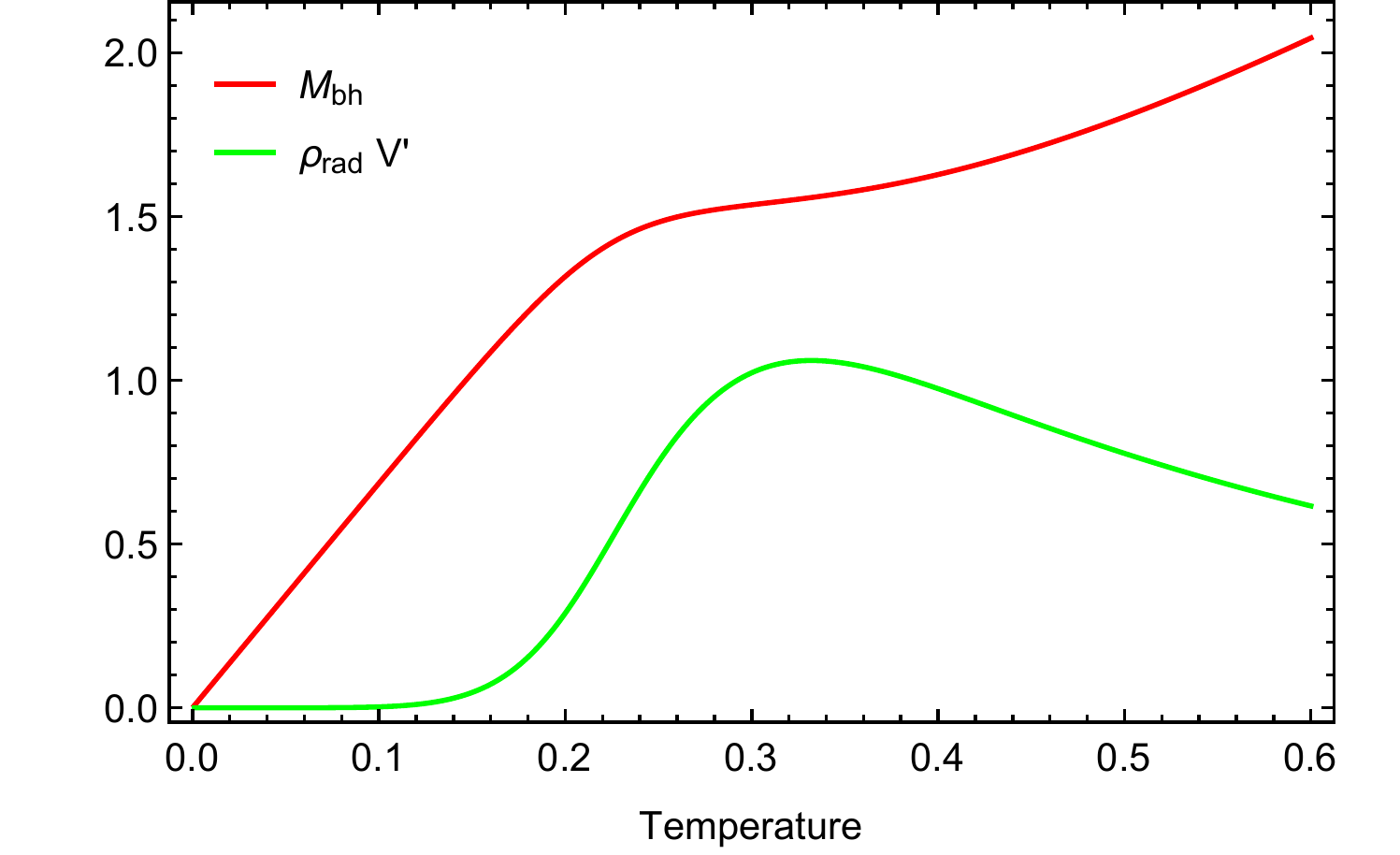}
\caption{Here we plot  $M_{bh}$ and $\rho_{rad} V' $ (where $V'= \frac{4 \pi}{3} R'^3$)as a function of temperature to check the validity of approximation. As we can see, $\rho_{rad} V'$  is always much smaller than  $M_{bh}$ at all temperatures, which means that the background energy density of the spacetime can be ignored safely.}
\label{approx}
\end{center}
\end{figure}

To check our approximation, we plotted  $M_{bh}$ and $\frac{4 \pi}{3} R'^3 \rho_{rad} $ as a function of temperature in Fig.~(\ref{approx}). It is evident from the plot that  $M_{bh}$ is always significantly larger than the $\rho_{rad} V'$ at all temperatures, hence justifying our approximation.

We note here that the exact black hole solutions in expanding spacetimes are most likely described by a wide class of solution that the so-called McVittie solution belongs to  \cite{McVittie:1933zz,Faraoni:2013aba,Kaloper:2010ec,Guariento:2012ri,Afshordi:2014qaa}. These solutions take into account the whole contribution from the MSH mass, however, no
single perfect fluid description can be used as a source for these solutions, which makes their straightforward interpretation difficult.


\end{document}